\documentclass[a4paper,12pt]{article}
\usepackage[utf8x]{inputenc}
\usepackage{graphicx,subfigure,color}
\usepackage{latexsym,amsmath,amssymb,amsfonts,amsthm,enumerate}
\usepackage{cite}

\newcommand{\e}[2][def]{
\begin{equation}
#2 \label{#1}
\end{equation} }

\begin{document}
\thispagestyle{empty}

\vspace*{1cm}

\begin{center}
{\large{\bf Complex Time Evolution of Open Quantum Systems.\\}} \vspace{1.8cm}
{\large C.~N. Gagatsos, A.~I.~Karanikas and G.~I.~Kordas}\\
\smallskip
{\it University of Athens, Physics Department\\
Nuclear \& Particle Physics Section\\
Panepistimiopolis, Ilissia GR 15771, Athens, Greece}\\
\vspace{1cm}
\end{center}
\vspace{0.4cm}

\begin{abstract}
We combine, in a single set-up, the complex time parametrization in path integration, and the closed time formalism of non-equilibrium field 
theories to produce a compact representation of the time evolution of the reduced density matrix. In this framework we introduce a cluster-type 
expansion that facilitates perturbative and non-petrurbative calculations in the realm of open quantum systems. The technical details of some very 
simple examples are discussed.
\end{abstract}

\newpage

\vspace*{.5cm}

{\bf 1. Introduction.}

\vspace*{.5cm}

In recent years there has been increasing interest in the consistent description of the dynamics of
open quantum systems~\cite{petr,wei,gjksz,dav,exn}. Quantum decoherence and dissipation are very important phenomena in many
different areas of physics. A non-exhaustive list includes problems from quantum optics to many body and
field-theoretical systems. Dissipative processes play a basic role in the quantum theory of lasers and photon
detection, and they are equally important in nuclear fission and the deep inelastic collisions of heavy ions.
More recently, the influence of the environment on a quantum system emerged as an issue of crucial
importance, not only due to its fundamental implications, but also due to its practical applications in
quantum information theory~\cite{pres,nie,alick}. In fact, during the last decade, many new discoveries regarding the
physics of open quantum systems were made. Primary examples of a promising progress can be found in
the rapidly developing field of quantum optics and the connected continuous variable systems in quantum
computation~\cite{cerf,brau}.

Theoretical studies of decoherence and dissipation in quantum mechanics are centered
on the time evolution of the reduced density matrix of a system embedded in a specific environment. The
basic tools for studying the reduced dynamics are either effective equations of motion, where the
dynamics of the environment are eliminated, such as the Lindblad master equation~\cite{lind,lidar}, or the influence
functional technique introduced by Feynman and Vernon~\cite{fey}. The latter is based on the path integral approach,
and was used by A. Caldeira and A. Leggett~\cite{cale} in the study of the quantum Brownian motion more than
twenty years ago. In most cases, however, neither the Lindblad equation nor the influence functional can be
exactly evaluated, since the interaction between the system and the environment is too complicated.
In fact, the simulation
of the environment by a system whose degrees of freedom are treated as random
variables following a more or less simple distribution, is a rather common practice. Therefore, one usually relies on
some simple, specific system-environment models: a harmonic oscillator or a two-level
quantum mechanical system embedded in a (thermal) bath of other harmonic oscillators
or other spin systems. In the present work we aim to introduce and investigate calculational
tools capable of exploring the behavior of an open system in interaction with a specific
quantum environment.
To be precise, we investigate the possibility
to extend the calculational capability of the Feynman-Vernon path integral approach by adopting and
combining definite functional methodological tools already known from different research fields. The first
such tool is a combination of the well-known “closed (real) time formalism”~\cite{chsuhayu} with the (equally well-known)
imaginary time formulation~\cite{wei} in the context of path integration. The compound result, called “closed complex
time formalism” (or CCT ), enables us to isolate, in a simple and compact expression, the influence of the
environment on the evolution of the system. 
It is well known that, in general, the integration of the environmental degrees
of freedom does not produce a local “effective action” that controls the
dynamics of the sub-system. The so-called Feynman-Vernon action, which incorporates the
influence of the environment, is a highly non-local object: it is non-local in time and in
space. The proposed CCT technique has a well-defined result: it produces an influence
functional that can be viewed as an action local in space. In this action the paths are
defined on the complex plane and they are parametrized with the help of a “time” running
on a specific contour of the complex plane. The interest in such formalism is not
“theoretical” but practical: one hopes to transfer the existent richness of perturbative and
non-perturbative path integral techniques into the realm of open quantum systems.

Our second suggestion, strongly related to the first one, is the application of the so-called
“cluster expansion” in the CCT context. The foundation of the application of
this very powerful technique, of course, lies in the spatial locality -on the complex plane- of the
influence functional. The cluster -or cumulant- expansion results to an expression that can
be viewed as the “effective action” that governs the dynamics of the system after the
elimination of the environmental degrees of freedom. However, in general, the cluster
expansion produces an infinite series that contains all the orders of the environmental
connected correlators and, if it is to be useful, some kind of truncation is necessary. As a
first step in this direction, we consider the case in which the environmental correlators are of
very fast decrease. Our formalism allows us to prove quite
generally and without any reference to a specific model, that the two-point environmental
correlator (which is the most important in our approximation scheme) has all the
properties that can lead the subsystem to decoherence and dissipation.
 It is worth noting that our proposal can be extended to systems with an
infinite number of degrees of freedom, such as the electromagnetic field interacting with matter or other
field-theoretical systems.

The remainder of the paper is organized as follows: In Section 2 we present the details of the
complex time formalism in the context of the path integral formulation of the Feynman-Vernon influence
functional, and we discuss the assumptions under which the aforementioned formalism is applicable. In
Section 3 we apply the cluster expansion in the framework of the CCT formalism, and we discuss the
emergence of some quite general and very important properties of the influence functional. In subsection
3.1 we provide a specific example of an environment which is just a simple harmonic oscillator (or a
collection of non-interacting harmonic oscillators). In Section 4 we consider the case of an environment in
which the correlations decay very fast after some characteristic time interval. This stochastic behavior
truncates the cluster series, enabling explicit calculations pertaining to the open system per se. As a first
step in this direction, in the same section we calculate the entanglement entropy of a simple
harmonic oscillator. Finally, in Appendix A we present the details of the calculation needed for deriving the
results appearing in section 4.

\vspace*{.5cm}

{\bf 2. Time Evolution and the Closed Complex Time Formalism.}

\vspace*{.5cm}

The best way to interpret the usefulness of the closed complex time methodology (CCT from now on) is the examination of 
the time evolution of the reduced (environment averaged) density matrix of an open quantum central system
(s from now on), which interacts linearly with its environment (e from now on).  
Adopting the usual starting point we assume that the total Hamiltonian can be written as the sum of two parts that refer to the system and the environment respectively, and a third part describing their interaction:
\e[01]{\hat{H}=\hat{H}_s+\hat{H}_e+\hat{H}_I}
The total system evolves in time unitarily and, consequently, the reduced density matrix changes in time 
according to the equation:
\e[02]{\hat{\rho}^R(t,t_0)=tr_e[\hat{U}(t,t_0)\hat{\rho}(t_0)\hat{U}^{\dag}(t,t_0)]}
The dynamical content of the last expression is incorporated into a time evolution operator that contains the degrees of 
freedom of the whole system:
\e[03]{\hat{U}(t,t_0)=\hat{T}\exp \left\{-\frac{i}{\hbar}\int_{t_0}^tdt'\hat{H}(t')\right\}}
In the last expression we have taken into account a possible time dependence of the Hamiltonian. 
Physically we understand such dependence in various ways; for example, we can imagine that, after a sudden quench, the coupling between the central system and its environment changes to a different value, remaining constant henceforth.  A case of physical interest arises when the coupling changes continuously and slowly enough to consider the evolution of the whole system as adiabatic.  Another example is the 
well-studied case of an external time dependent field coupled linearly to the central system. In any case the operator $\hat{T}$
takes care of the needed time ordering. For now let us assume that, at the initial time $t_0$ (for the sake of convenience, in what follows 
we shall assume that $t_0=0$), the total system is prepared in a pure disentangled state
\e[04]{|\psi\rangle = |\psi_s\rangle \otimes |\psi_e\rangle.}
Consequently, we can rewrite the reduced density matrix in the form:
\e[05]{\hat{\rho}^R(t)=tr_e[\hat{U}(t)\hat{\rho}_s(0)\otimes\hat{\rho}_e(0)\hat{U}^{\dag}(t)]}
Denoting by $\vec{x}$ and $\vec{q}$ the coordinates of the central system and the environment respectively, and by $\vec{X}=(x_1,...,x_D;q_1,...,q_D)$
the coordinates of the whole system collectively, eq. (\ref{05}) can be written in the well known form:
\e[06]{\rho^R_{\vec{x}'\vec{x}}(t)=\int d^D x'' \int d^D x''' \rho^s_{\vec{x}''\vec{x}'''}(0) J(\vec{x},\vec{x}',\vec{x}'',\vec{x}''';t)}
where the propagating kernel can be read from the expression:
\e[07]{J(\vec{x},\vec{x}',\vec{x}'',\vec{x}''';t)\equiv \int d^D q \int d^D q'' \int d^D q''' \rho^e_{\vec{q}''\vec{q}'''}(0) \langle \vec{x}',\vec{q}|\hat{U}(t)|\vec{X}''\rangle\langle\vec{X}'''|\hat{U}^{\dag}(t)|\vec{X}\rangle.}
Our next assumption is that the environment is initially in its ground state:
\e[08]{|\psi_e\rangle=|0_e\rangle}
Then, it can easily be shown that~\cite{cawil}:
$$\rho^e_{\vec{q}'',\vec{q}'''}(0)=\langle \vec{q}''|0_e\rangle\langle0_e|\vec{q}'''\rangle=$$
\e[09]{=\frac{1}{Z_e}\mathop{\int \mathcal{D}\vec{q}^{(3)}}\limits_{\vec{q}^{(3)}(-0)=\vec{q}''}  \mathop{\int\mathcal{D}\vec{q}^{(2)}}
\limits_{\vec{q}^{(2)}(+0)=\vec{q}''}  \exp\Bigg\{-\frac{1}{\hbar} \int_{-\infty}^{-0} d\tau \mathcal{L}_e^{(E)}[q^{(3)}]-
\frac{1}{\hbar}\int_0^{\infty} d\tau \mathcal{L}_e^{(E)}[q^{(2)}]\Bigg\}}
In the last expression we denoted by $\mathcal{L}_e^{(E)}$ the Euclidean version of the Lagrangian describing the dynamics of the environment. 
The origin of eq.(\ref{09}) can be traced back to the propagator:
$$G_e(\vec{q}'',t';\vec{q}''',t) = \sum_{n_{e}} \langle \vec{q}''|\exp\left\{-\frac{i}{\hbar}(t'-t)\hat{H}_e\right\}|n_e\rangle\langle n_e|\vec{q}'''\rangle=$$
\e[10]{=\sum_{n_{e}} \exp\left\{-\frac{i}{\hbar}(t'-t)E_{n_{e}}\right\} \phi_{n_e}(\vec{q}'')\phi_{n_{e}}^{\ast}(\vec{q}''')}
Introducing the Euclidean time $\tau=it$, taking the limits, $\tau=-T_E$, $\tau'=0$, $T_E\rightarrow \infty$ and assuming that the
ground state is unique one can easily deduce that
\e[11]{G_e(\vec{q}'',0;\vec{q}''',-T_E)=\langle \vec{q}''|0_e\rangle\langle 0_e|\vec{q}'''\rangle e^{-T_E E_{0_e}/\hbar}\Bigg[1+\mathcal{O}\left(e^{-T_E (E_{n_e}-E_{0_e})/\hbar}\right)\Bigg]}
and, consequently the ground state wave function can be determined through an integration of the Euclidean propagator:
\e[12]{\langle \vec{q}''|0_e\rangle \sim \int d\vec{q}''' G_e(\vec{q}'',0;\vec{q}''',-\infty)=\mathop{\int \mathcal{D}\vec{q}}\limits_{\vec{q}(-0)=\vec{q}''} \exp\left\{-\frac{1}{\hbar} \int_{-\infty}^{-0} d\tau \mathcal{L}_e^{(E)}[q]\right\}}
The above relation is the basis of eq.(\ref{09}) in which we also introduced the normalization factor
\e[13]{Z_e=\mathop{\int \mathcal{D}\vec{q}}\limits_{\vec{q}(-\infty)=\vec{q}(\infty)} \exp\left\{-\frac{1}{\hbar}\int_{-\infty}^{+\infty} d\tau \mathcal{L}_e^{(E)}[q]\right\}}
ensuring that $tr_e[\hat{\rho}_e(0)]=1$ and we used a numbering convenient for our future considerations. 
To proceed further we write:
\e[14]{\langle \vec{x}',\vec{q}|\hat{U}(t)|\vec{X}''\rangle = \mathop{\int \mathcal{D}\vec{X}^{(4)}}\limits_{\vec{x}^{(4)} (0)=\vec{x}''}^{\vec{x}^{(4)}(t)=\vec{x}'} \delta[\vec{q}^{(4)}(t)-\vec{q}] \delta[\vec{q}^{(4)}(0)-\vec{q}''] \exp \Bigg\{\frac{i}{\hbar}\int_{0}^{t} dt' \mathcal{L}_e[\vec{X}^{(4)}]\Bigg\}}
and
\e[15]{\langle\vec{X}'''|\hat{U}(t)|\vec{X}\rangle = \mathop{\int \mathcal{D}\vec{X}^{(1)}}\limits_{\vec{x}^{(1)} (0)=\vec{x}'''}^{\vec{x}^{(1)}(t)=\vec{x}} \delta[\vec{q}^{(1)}(t)-\vec{q}] \delta[\vec{q}^{(4)}(0)-\vec{q}'''] \exp \left\{\frac{i}{\hbar}\int_{t}^{0} dt' \mathcal{L}_e[\vec{X}^{(1)}]\right\}.}
Inserting eqs.(\ref{09}), (\ref{14}) and (\ref{15}) in expression (\ref{07}) we find:
$$J(\vec{x},\vec{x}',\vec{x}'',\vec{x}''';t)=$$
\e[16]{=\frac{1}{Z_e}\mathop{\int \mathcal{D}\vec{x}^{(1)}}\limits_{\vec{x}^{(1)}(0)=\vec{x}'''}^{\vec{x}^{(1)}(t)=\vec{x}} \mathop{\int \mathcal{D}\vec{x}^{(4)}}\limits_{\vec{x}^{(4)}(0)=\vec{x}''}^{\vec{x}^{(4)}(t)=\vec{x}'} 
\exp \Bigg\{\frac{i}{\hbar} \int_{0}^{t} dt' \mathcal{L}_s[x^{(4)}]+\frac{i}{\hbar} \int_{t}^{0} dt' \mathcal{L}_{s}[x^{(1)}]\Bigg\} \mathcal{F}[x^{(4)},x^{(1)};t].}
The last factor in the above equation defines the well-known Feynman-Vernon functional~\cite{fey} which incorporates the influence of the environment to the time evolution of the system:
$$\mathcal{F}[x^{(4)},x^{(1)};t]=\left(\prod_{i=1}^{4}\int \mathcal{D}\vec{q}^{(i)}\right) \delta[\vec{q}^{(4)}(0)-\vec{q}^{(3)}(-0)] \delta[\vec{q}^{(4)}(t)-\vec{q}^{(1)}(t)] \delta[\vec{q}^{(2)}(+0)-\vec{q}^{(1)}(0)]\times$$
\e[17]{\times \exp \Bigg\{\frac{i}{\hbar}\int_{0}^{t} dt' \mathcal{L}_{e+I} [\vec{X}^{(4)}]-\frac{1}{\hbar}\int_{-\infty}^{-0} d\tau \mathcal{L}_e^{(E)}[q^{(3)}]-\frac{1}{\hbar}\int_{0}^{\infty} d\tau \mathcal{L}_e^{(E)}[q^{(2)}]+\frac{i}{\hbar}\int_{t}^{0} dt' \mathcal{L}_{e+I} [\vec{X}^{(1)}]\Bigg\}}
Up to this point the only difference of the last result from the usual line of thinking~\cite{petr,wei,gjksz,dav} is that we consider the environment not as a heat bath in thermal 
equilibrium but as a quantum system -probably a very complicated one- in its ground state.
\begin{figure}
\centering
\includegraphics[width=6cm,height=7.8cm]{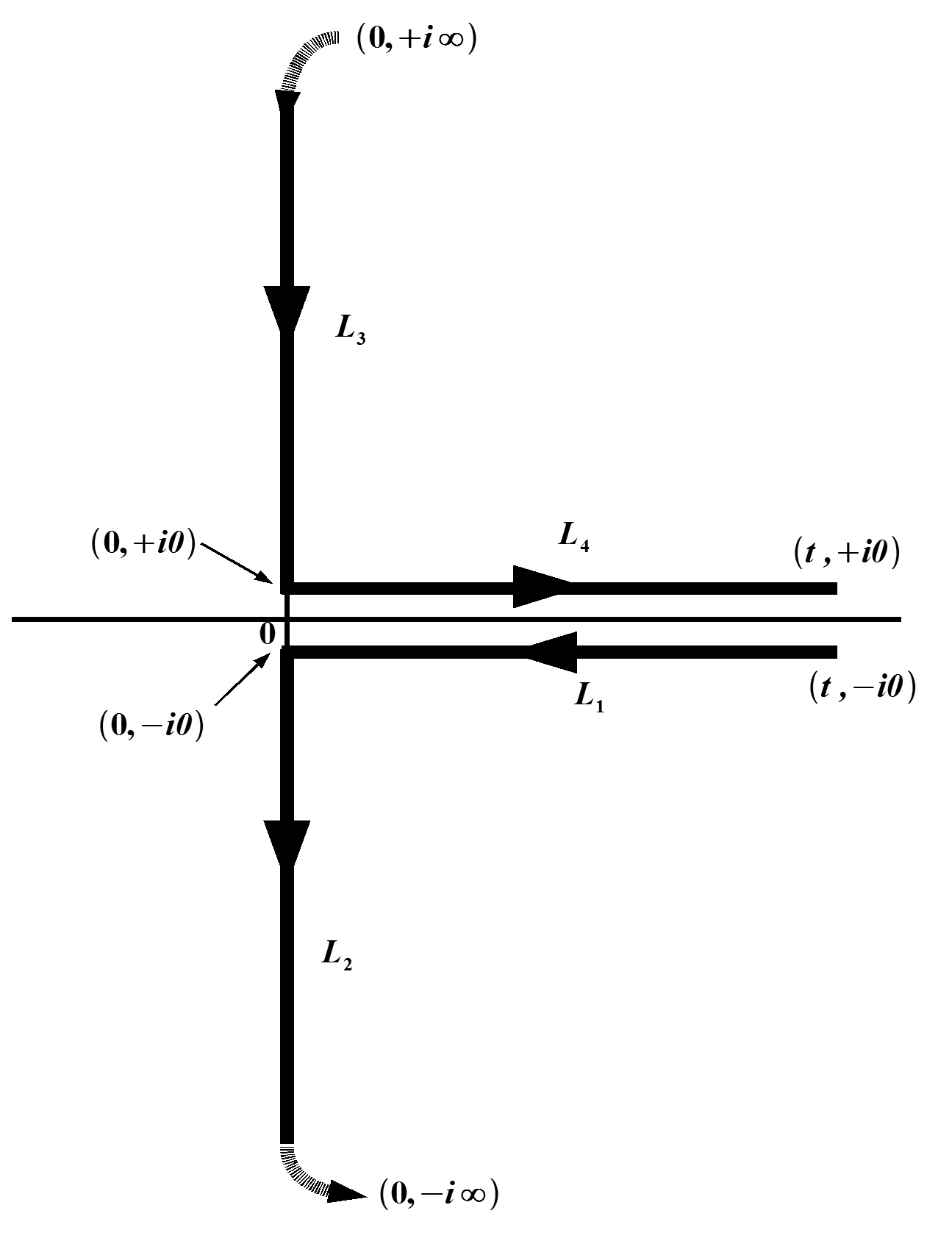}
\caption{The Contour C}
\end{figure}

The expression for the influence functional can now be considerably simplified if we introduce the complex variable $z$ defined on the contour $C$ 
shown in Fig. 1. This contour consists of 4 different straight lines: The first line $L_1$ goes parallel to the 
real axis from the point $\textit{z} = t -i0$ to point $\textit{z} = 0 -i0$. The second
line $L_2$ begins from the point $\textit{z} = 0 -i0$ and, following a path along the 
imaginary axis, goes to $\textit{z} = 0 -i \infty$. The line $L_3$ traces a path along the imaginary axis and joins the points $\textit{z} = 0 + i \infty$ and $\textit{z} = 0 + i0$. The last
part of the contour is the straight line $L_4$ : It goes parallel to the real axis from the point $\textit{z} = 0 + i0$ to
point $\textit{z} = t + i0$. It is now easy to be proved that the ``action'' in the influence functional (\ref{17}) can be written as follows:
\e[18]{ \tilde{S}=\int_{C} d\textit{z} \big\{\mathcal{L}_e[q_c]+\mathcal{L}_I[q_c,x_c]\big\}=
\int_{C} d\textit{z} \big\{\mathcal{L}_e[q_c]+g_c q_c x_c \big\}.}
The notation in the last equation is defined as follows: Along the lines $L_i$, $i=1,\ldots,4$ , we have
written $(\vec{x}_{L_i},\vec{q}_{L_i})=(\vec{x}^{(i)},\vec{q}^{(i)})$ and we have introduced a contour dependent coupling $g_c$ 
with $g_{L_{1}}=g_{L_{4}}=g$ and $g_{L_{2}}=g_{L_{3}}=0$. In expression (\ref{18}) we have explicitly assumed that the
interaction between the system and the environment is linear and has the minimal coupling $L_I = gxq$. In what
follows we shall also assume that the coupling $g$ is time independent, but our considerations can easily be
generalized to a time dependent coupling.

To confirm that eq.(\ref{18}) does indeed represent the action in the influence functional, 
let us note that along the lines $L_1$ and $L_4$ we can write $\textit{z}= t'-i0$, and consequently:
$$\int_{L_{1}}d\textit{z} \big\{\mathcal{L}_e[q_{L_{1}}]+\mathcal{L}_I[q_{L_{1}},x_{L_{1}}]\big\}=$$
\e[19]{\int_{t-i0}^{0-i0}d\textit{z}\big\{L_e[q_{L_{1}}]+\mathcal{L}_I[q_{L_{1}},x_{L_{1}}]\big\}=\int_{t}^{0}dt' L_{e+I}[\vec{X}^{(1)}]}
and
$$\int_{L_{4}}d\textit{z} \big\{\mathcal{L}_e[q_{L_{4}}]+\mathcal{L}_I[q_{L_{4}},x_{L_{4}}]\big\}=$$
\e[20]{\int_{0+i0}^{t+i0}d\textit{z}\big\{L_e[q_{L_{4}}]+\mathcal{L}_I[q_{L_{4}},x_{L_{4}}]\big\}=\int_{t}^{0}dt' L_{e+I}[\vec{X}^{(4)}]}
Along the lines $L_2$ and $L_3$ we write $z=0-i\tau$ and thus:
\e[21]{\int_{L_{2}}d\textit{z} \mathcal{L}_e [q_{L_{2}}]=\int_{0-i0}^{0-i\infty}d\textit{z} \mathcal{L}_e [q_{L_{2}}]=i \int_{0}^{\infty} d\tau \mathcal{L}_e^{(E)}[q^{(2)}]}
and
\e[22]{\int_{L_{3}}d\textit{z} \mathcal{L}_e [q_{L_{3}}]=\int^{0+i0}_{0+i\infty}d\textit{z} \mathcal{L}_e [q_{L_{3}}]=i \int^{-0}_{-\infty} d\tau \mathcal{L}_e^{(E)}[q^{(3)}]}
Inserting eqs.(\ref{19}), (\ref{20}), (\ref{21}) and (\ref{22}) into eq.(\ref{17}) and imposing the boundary condition
\e[23]{\vec{q}_{\textbf{\tiny C}}(-i \infty) = \vec{q}_{\textbf{\tiny C}}(+i\infty)}
we get the following compact expression for the influence functional:
$$\mathcal{F}[x^{(1)},x^{(4)};t] = \frac{1}{Z_e} \mathop{\int \mathcal{D}\vec{q}_{c}}\limits_{\vec{q}_{c}(t+i0)=\vec{q}_{c}(t-i0)}
\exp \Bigg\{\frac{i}{\hbar} \int_{c} d\textit{z} \mathcal{L}_e[q_{c}]+\frac{i}{\hbar}\int_{C}d\textit{z} g_{c}x_{c}q_{c}\Bigg\}\equiv$$
\e[24]{\equiv \left\langle\exp\left\{\frac{i}{\hbar}\int_c dz g_c q_c x_c \right\}\right\rangle_q \equiv \exp\left\{\frac{i}{\hbar} S_{FV}[x_c]\right\}}
($S_{FV}$ stands for the Feynman-Vernon action).

As it is obvious from the above expression, the introduction of the complex time $z$ defined on the contour $C$, has enabled us 
to interpret the influence functional as an integral over continuous paths with periodic boundary conditions.  
In fact, the compactness of the result indicated in eq. (\ref{24}) is the essence of the CCT formalism.  At this point it may be
useful to summarize our assumptions.  The first one was that initially the system and its environment were disentangled 
(see eq.(\ref{04})).  This assumption is not crucial either for the appearance of the influence functional or for the 
implementation of the closed complex time formalism. The basic assumption for the latter is that the time evolution begins 
from a ground state (see eq.(\ref{08})). To confirm this statement, let us assume that initially the central system and the 
environment were entangled and the whole was in the ground state of the Hamiltonian $H(\vec{P},\vec{X},g(0))$. Consider now 
the time evolution with a Hamiltonian $H(\vec{P},\vec{X},g(t))$ in which the coupling between the system and the environment 
changes very slowly (or has a different constant value). The evolution of the reduced density matrix reads now:
\e[25]{\rho^R_{\vec{x}'\vec{x}}(t)=\int d^D q \int d^{2D}X'' \int d^{2D}X''' \rho_{\vec{X}''\vec{X}'''}(0) \langle \vec{x}',\vec{q}| \hat{U}(t)|\vec{X}''\rangle
\langle \vec{X}'''|\hat{U}^\dagger (t)|\vec{X}\rangle.}
Following the reasoning that led us to eq. (9) we can write the initial density matrix in the form:
$$\rho_{\vec{X}''\vec{X}'''}(0)
 =\frac{1}{Z} \mathop{\int \mathcal{D}\vec{X}^{(3)}}\limits_{\vec{X}^{(3)}(-0)=\vec{X}''}
\mathop{\int \mathcal{D}\vec{X}^{(2)}}\limits_{\vec{X}^{(2)}(+0)=\vec{X}'''} \exp\Bigg\{-\frac{1}{\hbar} \int_{-\infty}^{-0}
d\tau \mathcal{L}^{(E)}[\vec{X}^{(3)},g(\tau)] - $$
\e[26]{ - \int^{\infty}_{0} d\tau \mathcal{L}^{(E)}[\vec{X}^{(2)},g(\tau)]\Bigg\}.}
Using now the expressions (\ref{14}) and (\ref{15}) we arrive at the result:
$$\rho^R_{\vec{x}'\vec{x}}(t)=\frac{1}{Z} \mathop{\int \mathcal{D}\vec{X}^{(4)}}\limits_{\vec{x}^{(4)}(t)=\vec{x}'} \int \mathcal{D}\vec{X}^{(3)}
\int \mathcal{D}\vec{X}^{(2)} \mathop{\int \mathcal{D}\vec{X}^{(1)}}\limits_{\vec{x}^{(1)}(t)=\vec{x}} \delta[\vec{X}^{(4)}(0)-\vec{X}^{(3)}(-0)]\times$$
$$ \times \delta[\vec{X}^{(2)}(+0)-\vec{X}^{(1)}(0)] \delta[\vec{q}^{(4)}(t)-\vec{q}^{(1)}(t)] \exp\Bigg\{\frac{i}{\hbar} \int_0^t dt'
\mathcal{L}[\vec{X}^{(4)}] - \frac{1}{\hbar} \int^{-0}_{-\infty} d\tau
\mathcal{L}^{(E)}[\vec{X}^{(3)}]-$$
\e[27]{- \int_{0}^{\infty} d\tau \mathcal{L}^{(E)}[\vec{X}^{(2)}] + \frac{i}{\hbar} \int^0_t dt'
\mathcal{L}[\vec{X}^{(1)}]\Bigg\}.}
Using the CCT formalism the last equation can be rewritten:
$$\rho^R_{\vec{x}'\vec{x}}(t)=\frac{1}{Z} \mathop{\int \mathcal{D}\vec{X}_c (z)}\limits_{\vec{x}_c(t-i0)=\vec{x}}^{\vec{x}_c(t+i0)=\vec{x}'} \exp\Bigg\{ 
\frac{i}{\hbar}\int_c dz \mathcal{L}[\vec{X}_c(z),g_c(z)]\Bigg\}=$$
\e[28]{=\frac{Z_e}{Z} \mathop{\int \mathcal{D}x_c (z)}\limits_{\vec{x}_c(t-i0)=\vec{x}}^{\vec{x}_c(t+i0)=\vec{x}'} \exp\Bigg\{ 
\frac{i}{\hbar}\int_c dz \mathcal{L}_s[x_c(z),g_c(z)]\Bigg\}\Bigg\langle \exp\Bigg\{ \frac{i}{\hbar} 
\int_c dz g_c q_c x_c\Bigg\}\Bigg\rangle_q}
Thus the evolution of the reduced density matrix assumes the compact form:
\e[29]{\rho^R_{\vec{x}'\vec{x}}(t)=\frac{Z_e}{Z} \mathop{\int \mathcal{D}\vec{x}_c (z)}\limits_{\vec{x}_c(t-i0)=\vec{x}}^{\vec{x}_c(t+i0)=\vec{x}'} \exp\Bigg\{ 
\frac{i}{\hbar}\int_c dz \mathcal{L}_s[x_c(z),g_c(z)] + \frac{i}{\hbar} S_{FV}\Bigg\}.}
A detailed analysis of the time evolution of a non-product initial state, under a time-dependent Hamiltonian, 
will be presented in a forthcoming study. For the time being we focus on the case of a disentangled initial state and a 
time-independent Hamiltonian. 

\vspace*{.5cm}

{\bf 3. The Cluster Expansion.}

\vspace*{.5cm}

It is self-evident that any further calculational step strongly depends on the dynamical details of the environment, as 
well as on the specific form of the interaction between the latter and the system. In any case, the compact formulation 
indicated in eqs.(\ref{24}) or (\ref{29}) can be combined with all the existent calculational technologies to produce concrete results 
in the field of open quantum systems. In this framework it is very convenient to use a well-known and very powerful technique:
The so-called cluster or cumulant expansion. This fundamental technique is widely used in a great variety of problems from 
statistical physics to quantum field theories~\cite{dosch}. The methodology has been extensively used in areas such as the resummation of 
perturbative series and non-perturbative estimations, among others, and has proven to be a very successful tool.

In our case, the cluster expansion theorem can be read from the relation:
$$\Bigg\langle \exp\Bigg\{ \frac{i}{\hbar} \int_c dz g_c q_c x_c\Bigg\}\Bigg\rangle_q = $$
\e[30]{= \exp\Bigg\{\sum_{n=1}^\infty \left(\frac{i}{\hbar}\right)^n \int_c dz_n...\int_c dz_1 \theta_c(z_n,...,z_1) K^{(n)}(z_n,...,z_1)\Bigg\}}
where
$$K^{(1)}(z_1) \equiv \langle\mathcal{L}_I [q_c(z_1)]\rangle_{q} = g_c(z_1) x_c(z_1) \langle q_c(z_1)\rangle_{q},$$
$$\phantom{0}$$
$$K^{(2)}(z_2,z_1) \equiv \langle\mathcal{L}_I [q_c(z_2)]\mathcal{L}_I [q_c(z_1)]\rangle_{q} - \langle\mathcal{L}_I [q_c(z_2)]\rangle_{q} \langle\mathcal{L}_I [q_c(z_1)]\rangle_{q}=$$
$$= g_c(z_2) g_c(z_1) x_c(z_2) x_c(z_1) \Big[\langle q_c(z_2)q_c(z_1)\rangle_{q} - \langle q_c(z_2)\rangle_{q}\langle q_c(z_1)\rangle_{q}\Big],$$
$$\phantom{0}$$
$$K^{(3)}(z_3,z_2,z_1) \equiv \langle\mathcal{L}_I [q_c(z_3)]\mathcal{L}_I [q_c(z_2)]\mathcal{L}_I [q_c(z_1)]\rangle_{q} - $$
$$- \Big\{\langle\mathcal{L}_I [q_c(z_3)]\rangle_{q} \langle\mathcal{L}_I [q_c(z_2)]\mathcal{L}_I [q_c(z_1)]\rangle_{q} + perms\Big\} + $$
$$ +\langle\mathcal{L}_I [q_c(z_3)]\langle\mathcal{L}_I [q_c(z_2)]\rangle_{q} \langle\mathcal{L}_I [q_c(z_1)]\rangle_{q},$$
\e[31]{.........etc.........}
In eq.(\ref{30}) we have introduced the chain of path dependent step functions
\e[32]{\theta_c (z_n,...,z_1) = \theta_c (z_n - z_{n-1})...\theta_c (z_2 - z_1),}
which takes care of the time ordering needed whenever the variables $z_i$ are integrated along the same line. The path dependent
step functions that appear in the above expression can be defined with the help of a proper parametrization of the contour
$z=z(\sigma),\phantom{0}\sigma\in [0,1]$ with $z(0)=t-i0$ and $z(1)=t+i0$. Since the time (real or Euclidean) flow follows 
different directions along different lines, we have introduced the following definition
\e[33]{\theta_c(z-z') = \theta_c(z(\sigma)-z(\sigma'))=\left\{\begin{array}{c}
                                                 \theta(\sigma-\sigma'),\phantom{0}when \phantom{0}C=L_{2,4} \phantom{0}.\\
                                                 \theta(\sigma'-\sigma),\phantom{0}when \phantom{0}C=L_{1,3}
                                               \end{array}\right.
}
When the variables $z$ are integrated along different lines the step functions become identically 1 or 0: For example, if $z\in L_1$
and $z'\in L_4$ we define $\theta_{L_1\cup L_4}(z-z')=1$ because the time along the line $L_1$ decreases, and this happens 
after its growth along the line $L_4$. 

The validity of eq.(\ref{30}) with the definition (\ref{33}) can be readily proven by expanding the corresponding exponentials. 
The proof can also be easily extended to the case of non-commutating quadratic matrices with the help of a proper time 
ordering. Taking into account the above conventions, any well-known result of the ordinary path integration can be 
transferred into the complex time framework as it is defined by the expressions (\ref{24}), (\ref{30}) and (\ref{31}).

From the preceding analysis we saw that the influence of the environment has been incorporated into the correlators
\e[34]{(i\hbar)^{n-1}\Delta^{(n)}_{C;i_n...i_1}(z_n,...,z_1)\equiv\langle q_{c,i_n}(z_n)...q_{c,i_1}(z_1)\rangle_{q}}
that must be integrated along a closed contour $C$ defined on the complex plane and  consisting of 4 lines in a definite order 
determined by the defining expression (\ref{02}) for the evolution of the density matrix. The time flow along the aforementioned 
contour is not causal, in the sense that its growth (along the line $L_4$) comes after its decrease (along the line $L_1$), 
a fact being taken into account in the properly defined path dependent step functions.
  
As it is evident from the definition of the path integral in eq.(\ref{17}), and the fact that the couplings disappear along the 
imaginary axis, non trivial correlations can exist only along the lines $L_1$ and $L_4$ or among them.  This is closely related 
to the fact that initially the central system and its environment were disentangled.
However, as we have already seen, the CCT formalism can also be applied if the system and its environment were initially 
entangled. In such a case, non trivial correlations can exist among all of the lines of the contour $C$.

At this point, we can highlight the properties of the fundamental functions (\ref{34}) by discussing some of the properties of 
the two point correlator which is supposed to be invariant under space rotations and time translations:
\e[34a]{\Delta^{(2)}_{C;i_2i_1}(z_2,z_1) \equiv \delta_{i_2i_1} G^{(2)}_c(z_2-z_1).}
A first observation is that it must have a non vanishing imaginary part due to the 
imaginary period over which it is defined. To be concrete, let us consider the propagation along the line $L_1$:
\e[35]{G_{L_1}^{(2)}(t_2-t_1)\equiv G_R(t_2-t_1)+iG_I(t_2-t_1).}
Along the line $L_4$ the time flow is reversed, and consequently:
\e[36]{G_{L_4}^{(2)}(t_2-t_1)\equiv G_R(t_1-t_2)+iG_I(t_1-t_2)=G_{L_1}^{(2)}(t_1-t_2).}
At this point we can appeal to the hermiticity of the density matrix: The influence functional must remain the same if 
we interchange $x^{(1)}$ and $x^{(4)}$ while taking the complex conjugate.  The last action reverses the time ordering along the 
contour $C$, and consequently the function $\Delta^{(2)}_{L_1}$ must be anti-hermitian:
\e[37]{[G_{L_1}^{(2)}(t_1-t_2)]^*=-G_{L_1}^{(2)}(t_2-t_1).}
Thus we immediately conclude that the real part of the propagator (\ref{35}) is an odd function, while its imaginary part is an even function of 
time:
\e[38]{G_R(t_2-t_1)=-G_R(t_1-t_2),\phantom{0}G_I(t_2-t_1)=G_I(t_1-t_2).}

The exchange contributions  can also be deduced with the same reasoning: Since, as we have discussed, the time along $L_1$
is after the time along $L_4$, the exchange from the line $L_4$ to the line $L_1$ is controlled by a function 
$G^{(2)}_{L_4\cup L_1}(t_2-t_1)$ in which $t_2< t_1$, while the exchange from the line $L_1$ to the line $L_4$ must be controlled 
by a function $G^{(2)}_{L_1\cup L_4}(t_2-t_1)$ in which $t_2> t_1$. Clearly the relation
\e[39]{G^{(2)}_{L_1\cup L_4}(t_2-t_1)=-[G^{(2)}_{L_4\cup L_1}(t_1-t_2)]^*}
must hold. The trace of the reduced density matrix must be equal to one, and, consequently, the Feynman-Vernon action must go to zero as
$x^{(4)}\rightarrow x^{(1)}$. This can happen only if the (forward) propagation $L_4 \rightarrow L_1$ exactly cancels the (forward) 
propagation along $L_4$, and the (backward) propagation $L_1 \rightarrow L_4$ exactly cancels the (backward) propagation along $L_1$:
\e[40]{G^{(2)}_{L_4\cup L_1}(t_2-t_1)=G_R(t_1-t_2)-iG_I(t_1-t_2)}
and
\e[41]{G^{(2)}_{L_1\cup L_4}(t_2-t_1)=-G_R(t_2-t_1)-iG_I(t_2-t_1).}
These arguments show clearly that, quite generally, the order $g^2$ contribution to the Feynman-Vernon action assumes the form:
$$\frac{i}{\hbar}S_{FV}^{(2)} = - \frac{1}{\hbar} \int^t_0 dt_2 \int^{t_2}_0 dt_1 [x^{(1)}(t_2)-x^{(4)}(t_2)]G_I(t_2-t_1)[x^{(1)}(t_1)-x^{(4)}(t_1)]+$$
\e[42]{+ \frac{i}{\hbar}\int^t_0 dt_2 \int^{t_2}_0 dt_1 [x^{(1)}(t_2)-x^{(4)}(t_2)]G_R(t_2-t_1)[x^{(1)}(t_1)+x^{(4)}(t_1)].}
It is now readily evident that the Feynman-Vernon action considerably changes the dynamics of the central quantum system. 
Its fluctuating part, which is connected to the imaginary part of the line propagator, reduces coherence. It is customary~\cite{petr,wei} and convenient to re-express its real part, 
which is connected to the real part of the line propagator, with the help 
of an even function $\gamma(t_2-t_1)=\gamma(t_1-t_2)$ through the relation:
\e[43]{G_R(t_2-t_1)=\frac{\partial}{\partial t_2} \gamma(t_2-t_1).}
The function $\gamma$ introduces in the Feynman-Vernon action a term which, on the classical level, can be understood as a damping 
or “friction”  term. Feeding eq.(\ref{42}) with expression (\ref{43}) we immediately find that:
$$\frac{i}{\hbar}S_{FV}^{(2)} =-\frac{1}{2\hbar} \int_0^t dt_2 \int_0^t dt_1 [x^{(1)}(t_2)-x^{(4)}(t_2)]G_I(t_2-t_1)[x^{(1)}(t_1)-x^{(4)}(t_1)]+$$
$$+\frac{i}{2\hbar} \int_0^t dt_2 \int_0^t dt_1 [x^{(1)}(t_2)-x^{(4)}(t_2)]\gamma(t_2-t_1)[\dot{x}^{(1)}(t_1)+\dot{x}^{(4)}(t_1)]-$$
$$-\frac{i}{\hbar}\gamma(0) \int_0^t dt_1 \Big[\big(x^{(1)}(t_1)\big)^2-\big(x^{(4)}(t_1)\big)^2\Big]+$$
\e[43a]{+\frac{i}{\hbar}[x^{(1)}(0)+x^{(4)}(0)] \int_0^t dt_1 [x^{(1)}(t_1)-x^{(4)}(t_1)]\gamma(t_1).}
At this point we must emphasize on the following observation: The last equation, being the exact
contribution of the second cumulant in the cluster expansion of the Feynman-Vernon action, is not an
approximate one. Despite the fact that it formally reproduces a colored -noise simulation of an
uncontrollable environment~\cite{petr,wei}, it is the first term in a systematic approximation of the environmental
dynamics.

\vspace*{.5cm}

{\bf 3.1. A Simple Example.}

\vspace*{.5cm}

As a specific example, let us compute, in the framework of the preceding analysis, the influence functional for the case in 
which the environment is just a simple harmonic oscillator
\e[44]{\mathcal{L}_e[\dot{\vec{q}},\vec{q}]=\frac{m_e}{2}\dot{\vec{q}}^{\phantom{|}2}-\frac{1}{2}m_e \omega_e^2 \vec{q}^{\phantom{|}2}.}
In this very simple case only one term appears in the rhs exponential in eq.(\ref{24}):
\e[45]{\frac{i}{\hbar} S_{FV} = - \frac{i}{\hbar}\int_C dz_2 g_c(z_2) \int_C dz_1 g_c(z_1) x_c(z_1)x_c(z_2) \theta_c (z_2 - z_1) G_C^{(2)} (z_2 - z_1).}
The Green function appearing in the last equation obeys periodic boundary conditions and assumes the well-known form
\e[46]{G_C^{(2)} (z_2 - z_1) = -\frac{1}{2m_e \omega_e}\frac{\cos[\omega_e(|z_2-z_1|_c-\tilde{T}/2)]}{\sin(\omega _e \tilde{T}/2)},}
with
\e[47]{|z_2-z_1|_c = (z_2-z_1)[\theta_c (z_2-z_1) - \theta_c (z_1-z_2)].}
The period is obviously imaginary $\tilde{T}=-2iT_E$, and consequently:
$$G_C^{(2)} (z_2 - z_1) \mathop{=}\limits_{T_E\rightarrow\infty}= -\frac{1}{2m_e \omega_e}[i\cos\omega_e|z_2-z_1|_c+\sin\omega_e|z_2-z_1|_c =$$
\e[48]{=-\frac{i}{2m_e \omega_e}e^{-i\omega_e|z_2-z_1|_c}.}
Given that $g_{L_1}=g_{L_4}=g$ and $g_{L_2}=g_{L_3}=0$ we can split the integration in (\ref{45}) as follows:
\e[49]{S_{FV}=g^2(I_{11} + I_{44} + I_{14}).}
Where we used the notation:
$$I_{11} = g^2\int_t^0 dt_2 \int _t^0 dt_1 \theta(t_1-t_2) x^{(1)}(t_2) x^{(1)}(t_1) \Bigg[i\frac{\cos\omega_e(t_2-t_1)}{2m_e \omega_e} + \frac{\sin\omega_e(t_2-t_1)}{2m_e \omega_e}\Bigg] = $$
\e[50]{ = \int^t_0 dt_2 \int^t_0 dt_1 \theta(t_2-t_1) y(t_2) y(t_1) [iG_I(t_2-t_1) + G_R(t_2-t_1)].}
In the last integral we have connected the result pertaining to the specific choice (\ref{44}) with the general result (\ref{35}) through 
the relations:
\e[51]{G_I(t_2-t_1)= \frac{g^2}{2m_e \omega_e} \cos\omega_e(t_2-t_1),\phantom{1}G_R(t_2-t_1)= -\frac{g^2}{2m_e \omega_e} \sin\omega_e(t_2-t_1).}
The second term in eq.(\ref{45}) reads:
$$I_{44} = g^2\int^t_0 dt_2 \int^t_0 dt_1 \theta(t_2-t_1) x^{(4)}(t_2) x^{(4)}(t_1) \Bigg[i\frac{\cos\omega_e(t_2-t_1)}{2m_e \omega_e} + \frac{\sin\omega_e(t_2-t_1)}{2m_e \omega_e}\Bigg] = $$
\e[52]{ = \int^t_0 dt_2 \int^t_0 dt_1 \theta(t_2-t_1) x^{(4)}(t_2) x^{(4)}(t_1) [iG_I(t_2-t_1) - G_R(t_2-t_1)].}
With the same reasoning the last term in eq.(\ref{45}) takes the form:
$$I_{14} = g^2 \int^t_0 dt_2 \int^t_0 dt_1 [\theta(t_2-t_1) + \theta(t_1-t_2)]x^{(4)}(t_2) x^{(1)}(t_1) \Bigg[-i\frac{\cos\omega_e(t_1-t_2)}{2m_e \omega_e} - \frac{\sin\omega_e(t_1-t_2)}{2m_e \omega_e}\Bigg] = $$
$$ = - \int^t_0 dt_2 \int^t_0 dt_1 \theta(t_2-t_1)x^{(4)}(t_2) x^{(1)}(t_1) [iG_I(t_2-t_1) + G_R(t_2-t_1)]-$$
\e[53]{- \int^t_0 dt_2 \int^t_0 dt_1 \theta(t_1-t_2)x^{(4)}(t_2) x^{(1)}(t_1) [iG_I(t_1-t_2) - G_R(t_1-t_2)].}
Inserting eqs.(\ref{50}), (\ref{52}) and (\ref{53}) into eq.(\ref{45}) we confirm the general result (\ref{42}) with 
the specific expressions (\ref{51}) for the  real and the imaginary part of the line propagator. These forms can be readily 
extended   to the case of a collection of $N$ harmonic oscillators:
$$g^2 G_I(t_2-t_1) \rightarrow \sum^N_{n=1}\frac{g_n^2}{2m_e \omega_{ne}}\cos\omega_{ne}(t_2-t_1)],$$
\e[54]{g^2G_R(t_2-t_1)\rightarrow -\sum^N_{n=1}\frac{g_n^2}{2m_e \omega_{ne}}\sin[\omega_{ne}(t_2-t_1)].}
The last expressions are obviously the $T\rightarrow 0$ limit of the well known result for an environment which is a heat bath 
consisting of a collection of harmonic oscillators in thermal equilibrium~\cite{cale}. 
\vspace*{.5cm}

{\bf 4. The Stochastic Environment.}

\vspace*{.5cm}

The cluster expansion discussed in the previous section, helped us to interpret the Feynman-Vernon action, and 
consequently the influence functional, as an infinite series over all possible correlations among the environmental degrees 
of freedom. However, it is evident, that such an interpretation can be useful only if the infinite series can be truncated 
with negligible error.  The case of weak coupling between the system and its environment is a first and obvious example; 
we shall not discuss this occurrence in the present paper, but is worth noting that the use of the cluster expansion 
facilitates the resummation of the (asymptotic) perturbative series. 

In the present study we adopt the hypothesis that the dynamics of the environment establish a characteristic time scale $\tau_e$
after which all internal correlations decay very fast:
\e[55]{G_I(t)=G_I\left(\frac{|t|}{\tau_e}\right) \longrightarrow 0,
\phantom{1}\gamma(t)=\gamma\left(\frac{|t|}{\tau_e}\right) \longrightarrow 0\phantom{1}for\phantom{1}|t|>\tau_e.}
The scale $\tau_e$ appearing in our starting relations (\ref{55}), is such a time interval that, when it elapses, the environment returns 
to its initial state. We shall also assume~\cite{petr,wei,pres,alick,kry} that there exists a second distinct time scale $\tau_s$, characterizing the interaction 
between the two parts of the entire system and, consequently, the evolution of the reduced density matrix, which is much 
larger than $\tau_e$: $\tau_s \gg \tau_e$.

In order to be more precise, let us assign an order of magnitude $||K^{(2)}||$ to the second order cumulant appearing in eq.(\ref{30}). 
We shall consider as stochastic the limit:
\e[56]{\frac{\tau_e}{\hbar} \sqrt{||K^{(2)}||}\rightarrow 0.}
As clearly shown by its definition $||K^{(2)}||$ is a measure of the average “strength” of the interaction between the central 
system and its environment: $\sqrt{||K^{(2)}||}\sim \langle V \rangle$. Defining the time scale $\tau_s$ as  $\tau_s \sim \hbar/\langle V \rangle$, 
the limit indicated in eq.(\ref{56}) can be obviously rephrased as $\tau_e/\tau_s \rightarrow0$.

We can now examine how the cluster expansion is formed at the stochastic limit. Assuming that $\langle q_c\rangle_{q}=0$ 
the first non-vanishing contribution comes from the second order term, which, following the discussion in the previous section,
assumes the quite general form (\ref{43a}).

As we are interested for $t\gg\tau_e$ we take into account eqs.(\ref{55}) and (\ref{56}), and, performing the expansion
$$x^{(i)}(t_2)\simeq x^{(i)}(t_1)+\mathcal{O}(t_2-t_1),$$
we get
\e[58]{g^2\int_0^t dt_2 [x^{(1)}(t_2)-x^{(4)}(t_2)]G_I(t_2-t_1) \approx \sigma [x^{(1)}(t_1)-x^{(4)}(t_1)].}
In the last expression we have introduced the quantity:
\e[59]{\sigma \equiv g^2\int_{-\infty}^{\infty} dt_2 G_I(t_2).}
In the same way, the second term in the rhs of eq.(\ref{43a}) can be approximated as follows:
\e[60]{g^2\int_0^t dt_2 [x^{(1)}(t_2)-x^{(4)}(t_2)]\gamma(t_2-t_1) \approx \lambda [x^{(1)}(t_1)-x^{(4)}(t_1)],}
with
\e[61]{\lambda \equiv g^2\int_{-\infty}^{\infty} dt_2 \gamma(t_2).}
With the help of a time rescaling $t_i=\tau_e \tilde{t}_i$ and using the defining relation for the $\gamma$ function 
(see eq. (43)) we can estimate that:
\e[62]{\frac{\lambda}{\sigma} \propto \tau_e \rightarrow 0.}
After the preceding approximations the second order contribution to the Feynman-Vernon action reads:
$$\frac{i}{\hbar}S^{(2)}_{FV} = - \frac{\sigma}{2\hbar} \int_0^t dt_1 [x^{(1)}(t_1) - x^{(4)}(t_1)]^2 + \frac{i\lambda}{2\hbar} \int_0^t dt_1 [x^{(1)}(t_1) - x^{(4)}(t_1)]
[\dot{x}^{(1)}(t_1) + \dot{x}^{(4)}(t_1)] -$$
\e[63]{- \frac{i}{\hbar} \gamma(0) \int_0^t dt_1 \Big[\big(x^{(1)}(t_1)\big)^2 - \big(x^{(4)}(t_1)\big)^2\Big] +
\frac{i\lambda}{2\hbar} \Big[\big(x^{(1)}(0)\big)^2 - \big(x^{(4)}(0)\big)^2\Big].}
Our claim is that, at the stochastic limit (\ref{56}), the cluster expansion, and consequently the Feynman-Vernon action, is 
dominated by the second order cumulant which, in this case, is expressed, by the above written eq.(\ref{63}). Indeed, each 
of the terms $K^{(n)}$ in the cumulant expansion represents a cluster that must be integrated over time intervals 
much larger than the time scale characterizing its exponential decay. Thus in the integrals
\e[64]{I^{(n)}=\int_0^t dt_n \int_0^{t_n} dt_{n-1} ... \int_0^{t_2} dt_1 K^{(n)}(t_n,...,t_1)}
the main contribution comes from time intervals $|t_1-t_i|\sim \tau_e$, $i=2,3,...$. Expanding the integrand as we have done in 
eqs.(\ref{58}) and (\ref{60}), we conclude that:
\e[65]{\frac{I^{(n)}}{I^{(n-1)}}=\mathcal{O}\left(\frac{\tau_e}{\tau_s}\right).}
This conclusion can be used to give concrete meaning to the environment characterized as stochastic:
It is the environment whose influence can be approximated by keeping only the second order correlator in the cluster 
expansion.

In other words, the Feynman-Vernon action, at the stochastic limit, can be approximated as follows:
\e[66]{S_{FV}\approx S_{FV}^{(2)}+\mathcal{O}\left(\frac{\tau_e}{\tau_s}\right).} 

At this point we must underline, once again, the strong resemblance of our result (\ref{63}) to the
case of the so-called Ohmic environment~\cite{petr,wei,gjksz,dav,cale}; that is, to the case of the quantum mechanical simulation of a
white-noise reservoir. Despite the fact that the expression (\ref{63}) for the Feynman-Vernon action is, in both
cases, formally the same, our result must be understood in a different context: It is the first term in a
systematic approximation of an exact result which is supposed to be valid at zero temperature. The
parameters appearing in eq.(\ref{63}) are not phenomenological, but they are strictly related to the two-point
correlation function of the environment, and, in principle, can be calculated at least numerically.
In the same context, the expression (\ref{24}) which is approximated by (\ref{66}), does not represent the
introduction of a random complex-valued Gaussian stochastic force: It is the specific environment under
consideration and its dynamics that justify the stochastic approximation. Having in mind the extension of
our work to infinite degrees of freedom, the non-Abelian gauge theories~\cite{giac} constitute the primary example of
such a stochastic behavior.

In the present study, the undertaken task is, so to speak, “phenomenological”: given the approximation (\ref{66}) for the influence of the 
environment, we try to estimate the consequences on the central system.

In any case, the result (\ref{66}) considerably facilitates the process of determining the time evolution of the reduced density 
matrix. The final result depends, of course, on the initial state of the central system, as well as on its specific dynamics. 
In what follows we shall consider the case in which the central system begins from its ground state
\e[67]{|\psi_s\rangle = |0_s\rangle.}
In such an occurrence we can use for $\rho^s(0)$ an expression analogous to the one (cf. eq.(\ref{09})) used in the previous section 
for the environmental density matrix:
$$\rho^s_{\vec{x}'' \vec{x}'''}(0)=
\frac{1}{Z_s} \mathop{\int\mathcal{D}\vec{x}^{(3)}}\limits_{\vec{x}^{(3)}(-0)=\vec{x}''} \mathop{\int\mathcal{D}\vec{x}^{(2)}}\limits_{\vec{x}^{(2)}(+0)=\vec{x}'''} \exp\Bigg\{-\frac{1}{\hbar}\int_{-\infty}^{-0} d\tau \mathcal{L}_{s}^{(E)}[x^{(3)}]\Bigg\}\times$$
\e[68]{\times\exp\Bigg\{-\frac{1}{\hbar}\int^{+\infty}_{+0} d\tau \mathcal{L}_{s}^{(E)}[x^{(2)}]\Bigg\}.}
Inserting the last expression into eq.(\ref{06}) we immediately get, at the stochasticity limit, the following path integral 
representation for the reduced density matrix:
\e[69]{\rho^R_{\vec{x}' \vec{x}}(t)\approx\frac{1}{Z_s Z_e}\mathop{\int \mathcal{D}\vec{x}_c(z)}\limits_{\vec{x}_c(t-i0)=\vec{x}}^{\vec{x}_c(t+i0)=\vec{x}'}\exp\Bigg\{\frac{i}{\hbar}
\int_c dz \mathcal{L}_s[x_c(z)]+\frac{i}{\hbar}S^{(2)}_{FV}[x_c(z)]\Bigg\}.}
As expressed in the last equation, the result for the reduced density matrix is simple and compact. This is due to the 
complex parametrization of the paths under integration. To obtain the final result, the integration over the central degrees 
of freedom must be performed and, obviously, this is a task that cannot be exactly accomplished in the general sense: some 
kind of approximation is needed. In any case 
eq.(\ref{69}) sets the scene where any available approximation technique can be performed. We can demonstrate the 
calculational abilities of our formalism by considering,once again, the zero order approximation i.e., the simple case in which the 
system is just one simple harmonic oscillator (we neglect any space index):
\e[70]{\mathcal{L}_s[x_c(z)]=\frac{1}{2}m \dot{x}_c^2-\frac{1}{2}m \omega^2 x_c^2.}
It is now suffices to observe that the contribution from the Feynman-Vernon action is quadratic, and consequently, the dependence 
of the reduced density matrix on the boundary values $x$ and $x'$ can be deduced just from the classical path:
\e[71]{m\left(\frac{d^2}{dz^2}+\omega^2\right)x_c^{cl.}(z)=\frac{\delta S_{FV}^{(2)}[x_c^{cl.}]}{\delta x_c^{cl.}(z)},\phantom{0}
x_c^{cl.}(t+i0)=x',x_c^{cl.}(t-i0)=x.}
In the last equation the rhs must be read in terms of the stochastic limit (\ref{63}). Thus we readily obtain:
$$\rho^R_{x' x}(t)\sim \exp\Bigg\{\frac{im}{2\hbar}[x' \dot{x}_c^{cl.}(t+i0)-x \dot{x}_c^{cl.}(t-i0)]\Bigg\}\times$$
\e[72]{\times\exp\Bigg\{-\frac{i}{2\hbar}\int_C dz x_c^{cl.}(z) \frac{\delta S_{FV}^{(2)}[x_c^{cl.}]}{\delta x_c^{cl.}(z)} + \frac{i}{\hbar} S_{FV}^{(2)}[x_c^{cl.}]\Bigg\}.}
The last two terms appearing in the rhs of the previous relation, cancel each other due to the quadratic nature of the truncated 
Feynman-Vernon action. Thus we conclude:
$$\rho^R_{x' x}(t)\sim \exp\Bigg\{\frac{im}{2\hbar}[x' \dot{x}_c^{cl.}(t+i0)-x \dot{x}_c^{cl.}(t-i0)]\Bigg\}=$$
\e[73]{=\exp\Bigg\{\frac{im}{2\hbar}[x' \dot{x}^{(4)}_{cl.}(t)-x \dot{x}^{(1)}_{cl.}(t)]\Bigg\}.}
The appearance of the classical trajectory in the last equation calls for the solution of the equation of motion (70). This is a 
lengthy but straightforward task, and it is presented in full detail in Appendix A. At this point it is enough to observe that 
the dependence of the classical solution on the boundary values $x$ and $x'$ is easily determined using the quite general ansartz:
\e[74]{\dot{x}_{cl.}^{(4)}(t)=\frac{1}{2}\alpha(t)x' + \frac{1}{2}\beta(t)x,\phantom{1}\dot{x}_{cl.}^{(1)}(t)=\frac{1}{2}\gamma(t)x' + \frac{1}{2}\delta(t)x.}
In the Appendix A we determine the coefficients in the above relations and confirm the validity of the relations $\delta(t)=\alpha^*(t)$ 
and $\gamma(t)=\beta^*(t)$, which are necessary for the hermiticity of the reduced density matrix. Inserting expressions (\ref{74}) 
in eq.(\ref{73}) we find that:
\e[75]{\rho^R_{x' x}(t) = C(t)\exp\Bigg\{\frac{im}{4\hbar}\Big[x'^2 \alpha(t) -x^2\alpha^*(t)+x'x\big(\beta(t)-\beta^*(t)\big)\Big]\Bigg\}.}
The suppression of the off-diagonal terms in the representation (\ref{75}) of the reduced density matrix is obviously related to the 
non-zero imaginary part of the function $\alpha(t)$, which in turn, as we confirm in the Appendix A, is related to the 
non-vanishing imaginary part of the environmental correlations. The normalization factor in equation (\ref{75}) is now determined 
by demanding:
\e[76]{C(t)\int_{-\infty}^\infty dx \exp\Bigg\{-\frac{m}{2\hbar}x^2 \Im\big(\alpha(t)+\beta(t)\big)\Bigg\}=1.}
The explicit calculations presented in Appendix A show that
\e[77]{\Im (\alpha(t)+\beta(t))=0}
yielding the conclusion that $C=1/L\rightarrow 0$, where $L$ is the volume of the space in which the system lives. In this 
case the reduced density matrix reads:
\e[78]{\rho^R_{x' x}(t) \sim \exp\left\{\frac{im}{4\hbar}\Re \alpha(t)(x'^2-x^2)\right\} \exp\left\{\frac{m}{4\hbar}\Im \alpha(t)(x'-x)^2\right\}.}
The explicit form of the function $\alpha(t)$ is presented in Appendix A. Here suffice it to note that $\Im \alpha$ is a positive definite 
increasing function of time. It is strictly related to the imaginary part of the environmental second order correlator 
since $\Im \alpha \varpropto \sigma$. Thus, the real factor of the density matrix (\ref{78}) is formally the density matrix of a 
free particle in a heat bath of temperature $k_B T = \frac{1}{2} \Im \alpha \varpropto \sigma$.

The exact time dependence of the function $\alpha(t)$ is tied with the value of the quantity:
\e[79]{q^2 = \frac{\lambda^2}{4m} + 2\frac{\gamma(0)}{m} - \omega^2.}
If $q^2>0$, $\alpha(t)$ becomes time independent for $t|q|\gg 1$ and
\e[80]{\Im \alpha \approx \frac{\sigma}{m}\frac{1}{|q|},\phantom{0} \Re \alpha\approx\frac{\lambda}{m}+2|q|.}
For $q^2=0$ and for $(\omega-\lambda /m)t\gg 1$, $\alpha(t)$ is again time independent:
\e[81]{\Im \alpha \approx \frac{\sigma}{m}\frac{2}{\omega-\lambda/m},\phantom{0} \Re \alpha \approx 2\omega.}
If $q^2\equiv-k^2<0$, and for $k t \gg1$, $\Im \alpha $ remains an increasing function of time:
\e[82]{\Im \alpha(t)\approx\frac{\sigma}{m}\frac{k^2+(\omega-\lambda/2m)^2}{[(\omega-\lambda/2m)\sin kt + k \cos kt ]^2}t.}

The reduced density matrix is the crucial quantity for the physics of an open system, playing a key role for determining 
all the system properties. As an interesting example, we shall focus on the entanglement entropy
\e[83]{S_{ent.}(t)=-tr_s\{\hat{\rho}^R(t)\ln \hat{\rho}^R(t)\}.}
The calculation of the entropy can be performed with the help of the so-called replica method~\cite{cawil}.
To apply it, one introduces the quantity
\e[84]{tr_s(\hat{\rho}^R)^n\equiv f(n)= \int dx^{(1)}\int dx^{(2)} ...\int dx^{(n)} \rho^R_{x^{(1)} x^{(2)}}\rho^R_{x^{(2)} x^{(3)}}...\rho^R_{x^{(n)} x^{(1)}}.}
After calculating the function $f(n)$ for integer $n$, we consider the function
\e[85]{f(\nu)=tr_s(\hat{\rho}^R)^\nu,\phantom{0} \nu>0.}
Using analytic continuation we can find the entanglement entropy from the relation
\e[86]{S_{ent.}=-\lim_{\nu\rightarrow1}\frac{tr_s(\hat{\rho}^R)^\nu-1}{\nu-1}=-tr_s\{\hat{\rho}^R(t)\ln \hat{\rho}^R(t)\}.}
Inserting eq.(\ref{78}) into eq.(\ref{84}) we get:
\e[87]{f(n)=C^n\left(\prod_{i=1}^n \int dx^{(i)}\right)\prod_{i=1}^n\exp\left\{-\frac{m}{4\hbar}\Im \alpha(t)(x^{(i+1)}-x^{(i)})^2\right\}, x^{(n+1)}=x^{(1)}.}
Consider now the propagation of a free particle with mass $m$ from the point $x$ to the point $x'$ in the Euclidean time interval
$t_E=2/\Im \alpha(t)$:
\e[88]{\mathop{\int\mathcal{D}x}\limits_{x(0)=x}^{x(t_E)=x'}\exp\left\{-\frac{m}{2} \mathop{\int_0 d\tau}\limits^{t_E}\dot{x}^2(\tau)\right\}
=\left[\frac{m\Im\alpha(t)}{4\pi\hbar}\right]^{1/2} \exp\left\{-\frac{m}{4\hbar}\Im \alpha(t) (x'^2-x^2)^2\right\}.}
Inserting the last expression into eq.(\ref{87}) we find that:
\e[89]{f(n)=\left[\frac{4\pi\hbar}{m\Im \alpha(t) L^2}\right]^{n/2} \mathop{\int\mathcal{D}x}\limits_{x(0)=x(nt_E)}\exp\left\{-\frac{m}{2} \mathop{\int_0 d\tau}\limits^{nt_E}\dot{x}^2(\tau)\right\}}
The last integral must be performed over periodic trajectories with period $nt_E$. Thus we can immediately conclude that:
\e[90]{f(n)=\left[\frac{4\pi\hbar}{m\Im \alpha(t) L^2}\right]^{n/2}\left[\frac{m\Im\alpha(t)L^2}{4\pi\hbar n}\right]^{1/2}.}
The entanglement entropy is now easily computed with the help of eq.(\ref{86}):
\e[91]{S_{ent.} = - \frac{1}{2} \ln \left[\frac{4\pi \hbar}{m\Im \alpha(t) L^2}\right]+\frac{1}{2}.}
It is worth noting that, as it is well-known, the entanglement entropy $S_{ent.}\sim \ln L$ is not an extensive quantity: 
contrary to the thermal entropy, is not analogous to the volume of the space in which the subsystem lives.  

\vspace*{.4cm}

{\bf 5. Conclusions and Perspectives.}

\vspace*{.4cm}

In this paper we have introduced two basic methodological tools for calculating the time evolution of the 
reduced density matrix and, consequently, the dynamics of an open quantum system. The first is the closed complex time 
(CCT) formalism, which combines two known approaches in a single set-up: The closed time formalism and the complex time one.
This formalism enabled us to express the time dependence of the reduced density matrix, in terms of a compact path integral, 
in which the paths are parametrized on a closed contour on the complex plane. Our second metodological suggestion is the introduction of 
the cluster expansion which is a very powerful tool, tested in a variety of problems, where the environmental details 
can be successfully approximated by keeping only the two-point correlators.  In this combined CCT-cluster expansion framework,
we examined the case of the so-called stochastic environment in which the correlations are decaying very “fast”.
In order to check our tools and examine the consequences of a stochastic environment, we performed a detailed “zero-order”
calculation for the simple case in which the system is a harmonic oscillator. We found the explicit form of the reduced 
density matrix as a function of time and we calculated the entanglement entropy. Depending on the details of the environment, 
the entropy is either a constantly increasing function of time or an increasing function of time that saturates to a constant
value. 

The purpose of this first work was to introduce and discuss the properties of a general
formalism that can be applied in a variety of problems. We have confined ourselves only to
a first -and in some sense trivial- application in order to demonstrate the underlying
calculational machinery. In a forthcoming study we shall present the far more interesting
case of the so-called quantum resonance. The general scene in such a problem is a double
well embedded in a stochastic environment and in interaction with an external time
dependent field. The path integral description of the tunneling and the role of the
“classical” solutions in the framework of CCT is a very interesting and far from trivial
problem that is under investigation.
\appendix
\setcounter{section}{0} \addtocounter{section}{1}
\section*{Appendix A}
\setcounter{equation}{0}
\renewcommand{\theequation}{\thesection.\arabic{equation}}

In this Appendix we shall determine the functions $\alpha(t)$ and $\beta(t)$ beginning from the classical equation of motion
\e[A1]{m\left(\frac{d^2}{dz^2}+\omega^2\right)x_c^{cl.}(z)=\frac{\delta S_{FV}^{(2)}[x_c^{cl.}]}{\delta x_c^{cl.}(z)}}
Due to its nonlocal character the above equation must be examined independently in every segment of the contour $C$.

Along the line $L_4$ the classical equation takes the form:
\e[A2]{m\left(\frac{d^2}{dt'^2}-\Omega^2\right)x^{(4)}_{cl.}(t')=i\sigma x^{(4)}_{cl.}(t')-\left(\lambda\frac{d}{dt'}+i\sigma\right)x^{(1)}_{cl.}(t'),}
where we defined
\e[A3]{m\Omega^2\equiv -m\omega^2+2\gamma(0).}
Along the lines $L_3$ and $L_2$ we have
\e[A4]{m\left(\frac{d^2}{d\tau^2}-\omega^2\right)x_{cl.}^{(3)}(\tau)=0}
and
\e[A5]{m\left(\frac{d^2}{d\tau^2}-\omega^2\right)x_{cl.}^{(2)}(\tau)=0.}
The last part of the classical equation refers to the line $L_1$:
\e[A6]{m\left(\frac{d^2}{dt'^2}-\Omega^{2}\right)x^{(1)}_{cl.}(t')=-i\sigma x^{(1)}_{cl.}(t')-\left(\lambda\frac{d}{dt'}-i\sigma\right)x^{(4)}_{cl.}(t').}
Seeking for continuous and differentiable solutions of the above system of classical equations, we impose the following 
boundary conditions:
\e[A7]{\begin{array}{c}
         x_{cl.}^{(4)}(t)=x',\phantom{0} x_{cl.}^{(4)}(0)=x_{cl.}^{(3)}(0)\\
         x_{cl.}^{(3)}(-\infty)=0 ,\phantom{0} x_{cl.}^{(3)}(0)=x_{cl.}^{(4)}(0)\\
         x_{cl.}^{(2)}(0)=x_{cl.}^{(1)}(0) ,\phantom{0}x_{cl.}^{(2)}(+\infty)=0\\
         x_{cl.}^{(1)}(t)=x ,\phantom{0}x_{cl.}^{(1)}(0)=x_{cl.}^{(2)}(0)
       \end{array}}
and
\e[A8]{\dot{x}_{cl.}^{(4)}(0)=\dot{x}_{cl.}^{(3)}(0),\phantom{0} \dot{x}_{cl.}^{(2)}(0)=-\dot{x}_{cl.}^{(1)}(0), \phantom{0} \dot{x}_{cl.}^{(3)}(-\infty)=\dot{x}_{cl.}^{(2)}(\infty).}
Equations (\ref{A4}) and (\ref{A5}) can be readily solved with the help of the above indicated boundary conditions:
\e[A9]{x_{cl.}^{(3)}(\tau)=x_{cl.}^{(4)}(0) e^{\omega\tau},\phantom{0} x_{cl.}^{(2)}(\tau)=x_{cl.}^{(1)}(0) e^{-\omega\tau}.}
Using once again the boundary conditions (\ref{A8}), we find that:
\e[A10]{\omega x_{cl.}^{(4)}(0)=\dot{x}_{cl.}^{(4)}(0),\phantom{0} \omega x_{cl.}^{(1)}(0)=\dot{x}_{cl.}^{(1)}(0).}
Introducing the combinations 
\e[A11]{y^{(\pm)}=\frac{1}{2}(x_{cl.}^{(4)} \pm x_{cl.}^{(1)}),}
the system of eqs.(\ref{A2}) and (\ref{A4}) can be considerably simplified:
\e[A12]{\left(\frac{d^2}{dt'^2}-\frac{\lambda}{m}\frac{d}{dt'}-\Omega^2\right)y^{(+)}(t')=2i\frac{\sigma}{m}y^{(-)}(t'),}
\e[A13]{\left(\frac{d^2}{dt'^2}-\frac{\lambda}{m}\frac{d}{dt'}-\Omega^2\right)y^{(-)}(t')=0.}
The solutions $y^{(\pm)}$ of the last equations are now trivially obtained and they lead us immediately to the result:
\e[A14]{x_{cl.}^{(4)}(t')=A_1 \varphi_+^{(4)}(t')+A_2 \varphi_-^{(4)}(t') + A_3 e^{\alpha_+t'} + A_4 e^{-\alpha_- t'},}
\e[A15]{x_{cl.}^{(1)}(t')=A_1 \varphi_+^{(1)}(t')+A_2 \varphi_-^{(1)}(t') + A_3 e^{\alpha_+t'} + A_4 e^{-\alpha_- t'}.}
In the above expression we have written:
\e[A16]{\varphi_+^{(4)}(t')= e^{\alpha_+t'} - 2i\frac{\sigma}{m}\int_0^t dt'' G(t',t'') e^{\alpha_+t''},}
\e[A17]{\varphi_-^{(4)}(t')= e^{-\alpha_-t'} - 2i\frac{\sigma}{m}\int_0^t dt'' G(t',t'') e^{-\alpha_-t''},}
\e[A18]{\varphi_+^{(1)}(t')= e^{\alpha_+t'} + 2i\frac{\sigma}{m}\int_0^t dt'' G(t',t'') e^{\alpha_+t''},}
\e[A19]{\varphi_-^{(1)}(t')= e^{-\alpha_-t'} + 2i\frac{\sigma}{m}\int_0^t dt'' G(t',t'') e^{-\alpha_-t''},}
\e[A20]{\alpha_\pm = \pm\frac{\lambda}{2m}+\sqrt{\frac{\lambda^2}{4m^2}+\frac{2\gamma(0)}{m}-\omega^2}.}
In eqs.(A.16) - (A.19) we used the Green’s function
\e[A21]{\left( \frac{d^2}{dt'^2}-\frac{\lambda}{m}\frac{d}{dt'}-\Omega^2\right)G(t',t'')=-\delta(t'-t''),\phantom{0} G(t,t'')=G(0,t'')=0}
which assumes the form:
$$G(t',t'')=\frac{e^{(\alpha_+ + \alpha_-)t/2-\alpha_+ t''}-e^{-(\alpha_+ + \alpha_-)t/2+\alpha_- t''}}{2(\alpha_+ + \alpha_-)\sinh[(\alpha_+ + \alpha_-)t/2]} (e^{\alpha_+ t'}-e^{-\alpha_- t'})\theta(t''-t')+$$
\e[A22]{+\frac{e^{(\alpha_+ + \alpha_-)t/2-\alpha_- t'}-e^{-(\alpha_+ + \alpha_-)t/2+\alpha_+ t'}}{2(\alpha_- + \alpha_+)\sinh[(\alpha_+ + \alpha_-)t/2]} (e^{\alpha_- t''}-e^{-\alpha_+ t''})\theta(t'-t'').}
The coefficients in eqs.(\ref{A14}) and (\ref{A15}) can straightforwardly be obtained with the help of the boundary conditions 
(\ref{A7}) and (\ref{A10}):
\e[A23]{A_1(t)=-\frac{\lambda_-(t)}{D(t)} \frac{x'-x}{2},}
\e[A24]{A_2(t)=\frac{\lambda_+(t)}{D(t)} \frac{x'-x}{2},}
\e[A25]{A_3(t)=\frac{\alpha_- + \omega}{\tilde{D}(t)} \frac{x'+x}{2}+\frac{\mu_+(t) \lambda_-(t) - \mu_-(t)\lambda_+(t)}{\tilde{D}(t)D(t)}e^{-\alpha_- t} \frac{x'-x}{2},}
\e[A26]{A_4(t)=\frac{\alpha_+ - \omega}{\tilde{D}(t)} \frac{x'+x}{2} - \frac{\mu_+(t) \lambda_-(t) - \mu_-(t)\lambda_+(t)}{\tilde{D}(t)D(t)}e^{\alpha_+ t} \frac{x'-x}{2},}
with
\e[A27]{D(t)=\lambda_+(t) e^{-\alpha_- t} - \lambda_-(t) e^{\alpha_+ t},
\tilde{D}(t)= (\alpha_+ - \omega) e^{-\alpha_- t} + (\alpha_- + \omega) e^{\alpha_+ t},}
\e[A28]{\lambda_\pm (t) = \dot{\varphi}_\pm^{(4)} (0) + \dot{\varphi}_\pm^{(1)} (0) - 2\omega,\phantom{0} \mu_\pm = \frac{1}{2}(\dot{\varphi}_\pm^{(4)}(0) - \dot{\varphi}_\pm^{(1)}(0)).}
Inserting (\ref{A22}) and (\ref{A23}) into (\ref{A14}) and (\ref{A15}), we determine:
$$\alpha(t)=\frac{\lambda_+(t)\dot{\varphi}_-^{(4)}(t)-\lambda_-(t)\dot{\varphi}_+^{(4)}(t)}{D(t)}+ (\alpha_+ - \alpha_-)\frac{\mu_+(t)\lambda_-(t) - \mu_-(t)\lambda_+(t)}{\tilde{D}(t)D(t)}e^{(\alpha_+-\alpha_-)t} +$$
\e[A29]{+\frac{\alpha_+(\alpha_-+\omega)e^{\alpha_+ t}-\alpha_-(\alpha_+-\omega)e^{-\alpha_- t}}{\tilde{D}(t)},}
$$\beta(t)=-\frac{\lambda_+(t)\dot{\varphi}_-^{(4)}(t)-\lambda_-(t)\dot{\varphi}_+^{(4)}(t)}{D(t)} - (\alpha_+ - \alpha_-)\frac{\mu_+(t)\lambda_-(t) - \mu_-(t)\lambda_+(t)}{\tilde{D}(t)D(t)}e^{(\alpha_+-\alpha_-)t} +$$
\e[A30]{+\frac{\alpha_+(\alpha_-+\omega)e^{\alpha_+ t}-\alpha_-(\alpha_+-\omega)e^{-\alpha_- t}}{\tilde{D}(t)},}
$$\gamma(t)=-\frac{\lambda_+(t)\dot{\varphi}_-^{(1)}(t)-\lambda_-(t)\dot{\varphi}_+^{(1)}(t)}{D(t)} + (\alpha_+ - \alpha_-)\frac{\mu_+(t)\lambda_-(t) - \mu_-(t)\lambda_+(t)}{\tilde{D}(t)D(t)}e^{(\alpha_+-\alpha_-)t} +$$
\e[A31]{+\frac{\alpha_+(\alpha_-+\omega)e^{\alpha_+ t}-\alpha_-(\alpha_+-\omega)e^{-\alpha_- t}}{\tilde{D}(t)},}
$$\delta(t)=\frac{\lambda_+(t)\dot{\varphi}_-^{(1)}(t)-\lambda_-(t)\dot{\varphi}_+^{(1)}(t)}{D(t)} - (\alpha_+ - \alpha_-)\frac{\mu_+(t)\lambda_-(t) - \mu_-(t)\lambda_+(t)}{\tilde{D}(t)D(t)}e^{(\alpha_+-\alpha_-)t} +$$
\e[A32]{+\frac{\alpha_+(\alpha_-+\omega)e^{\alpha_+ t}-\alpha_-(\alpha_+-\omega)e^{-\alpha_- t}}{\tilde{D}(t)}.}
(The argument in all the functions is the instant $t$.)

At this point we are ready to confirm some of the claims presented in the main text. We must distinguish two cases. The first is when:
\e[A33]{\frac{\lambda^2}{4m^2}\geq \omega^2 - 2\frac{\gamma(0)}{m}}
In such a case $\alpha_\pm$ are real and consequently $\varphi_\pm^{(4)}=(\varphi_\pm^{(1)})^*$. Observing that
$\lambda_\pm=\lambda_\pm^*$, $\mu_\pm=-\mu_\pm^*$ we immediately see that:
\e[A34]{\gamma^*=\beta,\phantom{0} \delta^*=\alpha}
and
\e[A35]{\Im \alpha(t) = - \Im \beta(t).}
When
\e[A36]{\frac{\lambda^2}{4m^2}< \omega^2 - 2\frac{\gamma(0)}{m}}
we observe that $\alpha_+=-\alpha_-^*$, $\varphi_\pm^{(4)}=(\varphi_\mp^{(1)})^*$, and since $\lambda_\pm,\mu_\pm$ turn out to be the 
same as in the case (\ref{A33}), we verify once again the relations (\ref{A34}) and (\ref{A35}).

When $\alpha_\pm$ are real we straightforwardly obtain:
\e[A37]{\Im \alpha(t) = \frac{\sigma}{m}\left[\frac{e^{(\alpha_+-\alpha_-)t/2}}{D(t)}f_1(t) + \frac{e^{(\alpha_+-\alpha_-)t}}{D^2(t)}f_2(t)\right],}
\e[A38]{\Re \alpha(t) = 2\frac{d}{dt}\ln D(t),}
with
$$f_1(t)=\frac{1}{\sinh[(\alpha_+ + \alpha_-)t/2]}\Bigg\{(\alpha_+ - \omega)\left[t-\frac{1-e^{-(\alpha_+ +\alpha_-)t}}{(\alpha_+ +\alpha_-)}\right] +$$
\e[A39]{+ (\alpha_- + \omega)\left[\frac{e^{(\alpha_+ +\alpha_-)t}-1}{(\alpha_+ +\alpha_-)}-t\right]\Bigg\}}
and
$$f_2(t)=\frac{\alpha_+ + \alpha_-}{\sinh[(\alpha_+ + \alpha_-)t/2]}\Bigg\{(\alpha_- + \omega)\left[te^{(\alpha_+ +\alpha_-)t/2}- 2\frac{\sinh[(\alpha_+ + \alpha_-)t/2]}{\alpha_+ + \alpha_-}\right]+$$
\e[A40]{+(\alpha_+ - \omega)\Bigg[2\frac{\sinh[(\alpha_+ + \alpha_-)t/2]}{\alpha_+ + \alpha_-}-te^{-(\alpha_+ +\alpha_-)t/2}\Bigg].}
The last relations confirm that $\Im\alpha>0$. For $t\alpha_\pm \gg 1$ it is easy to check that $\Im\alpha$ and $\Re\alpha$
become constants:
\e[A41]{\Im \alpha \approx \frac{2\sigma}{m}\frac{1}{\alpha_+ +\alpha_-}, \phantom{0} \Re \alpha \approx 2 \alpha_+.}
The last relation holds as long as $\alpha_+ +\alpha_- \neq 0$. If $\alpha_+ +\alpha_- = 0$ that is if
\e[A42]{\frac{\lambda^2}{4m^2}+ 2\frac{\gamma(0)}{m}=\omega^2 }
we immediately find that
\e[A43]{\Im \alpha = \frac{\sigma}{m}\frac{2t}{1+(\omega-\lambda/2m)}\mathop\approx\limits_{t\rightarrow\infty} \frac{\sigma}{m}\frac{2}{\omega-\lambda/m},}
\e[A44]{\Re \alpha = 2\frac{\omega+t(\omega-\lambda/2m)\lambda/2m}{1+t(\omega-\lambda/2m)}\mathop\approx\limits_{t\rightarrow\infty} 2\omega.}
When $\alpha_\pm$ are complex we find that:
$$\Im \alpha(t)=\frac{\sigma}{m}\frac{1}{[(\omega-\lambda/2m)\sin kt + k \cos kt ]^2}\Bigg\{\left[k^2+(\omega-\lambda/2m)^2\right]t+$$
\e[A45]{+ 2(\omega-\lambda/2m)\sin^2 kt + \left[k^2-(\omega-\lambda/2m)^2\right]\frac{\sin 2kt}{2k}\Bigg\}.}
Using the fact $x/\sin x \geq1$ once again we can verify that $\Im \alpha(t)> 0$. It also straightforward to see that:
\e[A46]{\Re \alpha(t)= 2\frac{d}{dt}\ln D(t)=2\frac{\left[\frac{\lambda}{2m}(\omega-\lambda/2m)-k^2\right]\sin kt +k\cos kt}{(\omega-\lambda/2m)\sin kt + k\cos kt},}
where we have noted
\e[A47]{k^2=\omega^2-\frac{2\gamma(0)}{m}-\frac{\lambda^2}{4m^2}.}

\end{document}